
\magnification=1200
\vsize=7.5in
\hsize=5in
\pageno=1
\tolerance=10000
\null
\vskip 1.0in
\centerline{\bf ABSTRACT}
\bigskip
\baselineskip 24pt plus 4pt minus 4pt
I study phase transitions occuring in noncollinear magnets by means of
a self-consistent screening approximation. The Ginzburg-Landau theory involves
two N-component vector fields with two independent quartic couplings allowing
a symmetry-breaking scheme which is $SO(N)\times SO(2) \rightarrow
SO(N-2)\times
SO(2)_{diag}$. I find that there is a second-order phase transition in the
physical
cases N=2,3, D=3 and that there is no fluctuation-induced first-order
transition. This is very similar to the case of the normal-to-superconducting
phase transition as recently found by Radzihovsky. The exponents are $\eta
(N=3,
D=3)\approx 0.11$, $\eta (N=2, D=3)\approx 0.15$ and go smoothly to the large-N
limit.
\vfill
\eject
\magnification=1200
\baselineskip 24pt plus 4pt minus 4pt
Many magnetic systems have a low-temperature ordered phase that breaks
completely
the rotation invariance due to a noncollinear pattern of the magnetic
moments$^1$.
Examples includes the rare-earths Ho, Dy and Tb. This is generic for
vector-spin
models with strong enough competing interactions. The phase transition
associated
with this ordering does not belong {\it a priori} to the well-studied
universality
classes corresponding to the collinear ordering with symmetry breaking
$O(N)\rightarrow O(N-1)$. Several groups have investigated the critical
behaviour
of the three-dimensional stacked triangular (STA) Heisenberg antiferromagnet
which is a simple example of commensurate noncollinear ordering and
there is general agreement$^{2,3,4}$ that this system has a second-order phase
transition with exponents$^{3}$
$\nu =0.585(9)$, $\gamma /\nu =2.011(14)$ that do not correspond to the $O(N)$
exponents. This peculiar set of exponents is apparently associated with a new
universality class: they appear also in the body-centered tetragonal
antiferromagnet$^5$. This class should also include the transition$^1$ from the
$^3$He liquid to the A-phase, Josephson-junction arrays in a transverse field
as
well as the fully frustrated bipartite lattice (Villain model). In the XY case
there is also a new set of exponents$^{1,2}$: $\nu=0.54(2)$, $\gamma =1.13(5)$

The renormalization group has been applied to the Ginzburg-Landau theory of
such
systems by Garel and Pfeuty$^6$. They used the expansion in $\epsilon =4-D$,
where D is the dimension, for any number N of components of the vector spins.
In
this framework, there is a Heisenberg $O(2N)$ fixed point which is always
unstable.
In the physical case N=3 there is no new fixed point  and this runaway is
interpreted as indicating a fluctuation-induced first-order phase transition.
This
is curiously similar to the normal-superconducting (NS) phase transition as
revealed by Halperin, Lubensky and Ma$^7$.

Recently, the NS phase transition has been studied by Radzihovsky$^8$ who used
the
self-consistent screening approximation (SCSA) due to Bray$^9$. This
approximation
includes an infinite subset of the 1/N expansion by means of a simple
self-consistency condition on the propagator. It allows to obtain the exponent
$\eta$ as a function of N and D. In the NS case the SCSA leads to a stable
fixed
point in the physical case D=3, in agreement with Monte-Carlo findings and
other
theoretical arguments$^{10}$. In the neighborhood of D=4, the corresponding
fixed
point survives for all values of the number of components of the order
parameter,
contrary to the prediction of the $4-\epsilon$ expansion, suggesting that the
fluctuation-induced first-order transition is an artifact of the $\epsilon$
expansion. It is the purpose of this Letter to apply
this very same method to the Ginzburg-Landau theory of helimagnets. I find that
there is a non-trivial fixed point that survives in the whole domain $2\le D
<4$ and
$2\le N \le\infty$ which is different from the O(2N) Heisenberg fixed point. In
the
physical cases of interest,  I find $\eta (N=3, D=3)\approx 0.11$ and $\eta
(N=2,
D=3)\approx 0.15$. This fixed point goes smoothly to the $N=\infty$ result and
there is no fluctuation-induced first-order phase transition. This is very
close to
the analysis of the NS transition of Ref.8. The mere existence of this fixed
point
for $N=2,3, D=3$ provides a natural explanation to the Monte-Carlo results and,
possibly, of some experimental results.

The Ginzburg-Landau theory for a generic Heisenberg helimagnet involves two
vector
fields that correspond to the Fourier modes of the magnetization near the
ordering
wavevectors $\pm {\vec Q}$. The effective action contains {\it two} quartic
invariants:
$$
A=
{1 \over 2}\left( {\left( {\nabla {\vec\phi} _1} \right)^2+\left( {\nabla
{\vec\phi} _2}\right)^2} \right)
+r\left( {{\vec\phi} _1^2+{\vec\phi} _2^2} \right)+
$$
$$
+u
\left( {\left( {{\vec\phi} _1^2+{\vec\phi} _2^2} \right)^2} \right)+v
 {\left({\left( {{\vec\phi} _1\cdot {\vec\phi} _2} \right)^2}
-{\vec\phi}_1^2 {\vec\phi}_2^2\right)}.
\eqno(1)
$$
This free energy has a global symmetry $O(N)\times O(2)$. When the coefficient
$v$
is positive, the ground state consists of orthogonal vectors and the residual
symmetry is $O(N-2)\times O(2)_{diag}$ where the subgroup $O(2)_{diag}$ acts
diagonally on vector indices and internal (1,2) indices$^{11}$.
The global symmetry allows only the two quartic invariants in Eq.(1).
A detailed study of this model has been performed by H. Kawamura$^{12}$.
In the
$D=4-\epsilon$ calculation, one finds the $O(2N)$ Wilson-Fisher fixed point on
the
line $v=0$ at a distance $\epsilon$ from the origin, in the ($u, v$) plane. The
operator $[{{\vec\phi} _1^2 {\vec\phi} _2^2-\left( {{\vec\phi} _1\cdot
{\vec\phi}
_2} \right)^2} ]$ opens a direction of instability for all values of N.
However, if
N is greater than $N_c(D)=21.8-23.4\epsilon +O(\epsilon^2)$ (obtained from a
two-loop computation$^{12}$), there is an additional stable fixed point with
$u^*\neq 0, v^*\neq 0$. In the neighborhood of the upper critical dimension,
there
is thus a dividing line
$N_c(D)$ in the (N, D) plane above which one has a second-order phase
transition
and below which one conjectures a fluctuation-induced first-order phase
transition.
This is similar to the normal  to superconductor transition$^{7,8}$ except for
numerical factors: there the transition is second-order only when the number of
complex components of the order parameter is larger than $\approx 183$.

The self-consistent screening approximation is an improvement over the large-N
expansion which has been applied with success to several physical
problems$^{8,9,13}$. In the leading large-N expression for the self-energy of
the
basic field containing the geometric sum of bubbles, one uses renormalized
propagators everywhere: this leads to a self-consistency equation that goes
beyond the simple result $N\rightarrow\infty$. An interesting byproduct is that
this approximation gives back the $N\rightarrow\infty$ result automatically.
Its
validity extends to all N and D values but, of course, this is not a systematic
procedure and we do not expect to get precise numerical values for the exponent
$\eta$. The propagator of the theory (1) is defined as $G_{11}(k)=\langle
\phi^\alpha_1 (k)\phi^\alpha_1 (-k)\rangle =\langle
\phi^\alpha_2 (k)\phi^\alpha_2 (-k)\rangle = G_{22}(k)\equiv G(k)$. The
self-energy is defined by $G^{-1}(k)=k^2 +r+\Sigma (k)$.
The Dyson equation is then:
$$
\Sigma (k) -\Sigma (0)=\int {d^D p\over (2\pi)^D} (G(k+p)-G(p))[4{\hat u}(p)+
{\hat v}(p)],
\eqno(2)
$$
where the free propagator is $G^{-1}_0 (k)=k^2 +r$ and the dressed vertices
${\hat u}(p)$, ${\hat v}(p)$ are given by series involving only powers of the
polarization bubble
$\Pi (k)=\int {d^D p\over (2\pi)^D} G(p)G(p+k)$:
$$
{\hat v}(p)={v\over 1+vN\Pi (p)}, \quad {\hat u}(p)={u+v(2u-v/2)N\Pi (p)\over
1+4uN\Pi (p)+v(4u-v)N^2 \Pi (p)^2}.
\eqno(3)
$$
Note that it is the full propagators that enter the quantities $\Pi (p), {\hat
u}
(p), {\hat v}(p)$ and thus Eq.(2) is a self-consistency condition. The
corresponding equation in the large-N limit is obtained by using the free
propagator $G_0$ in the right-hand side of Eq.(2).

The SCSA strategy$^9$ amounts to considering criticality i.e. $r=0$ and writing
the scaling form of the propagator $G^{-1}(k)=k_c^{\eta} k^{2-\eta}$ since
there
is no longer any correlation length. With this scaling form for $G$ one can
compute the polarization bubble following Bray$^9$:
$$
\Pi (p)= {\Gamma (2-D/2-\eta) \Gamma^2 (D/2+\eta/2-1)\over (4\pi)^{D/2} \Gamma
(D-2+\eta) \Gamma^2 (1-\eta/2)}\times k_c^{-2\eta} p^{D-4+2\eta}.
\eqno(4)
$$

The Heisenberg O(2N) fixed point can be found by setting $v=0$. If we suppose
that
$\Pi$ blows up for small momenta then ${\hat u}(p)\approx 1/4N\Pi$. One can now
match the powers of $k^{-2\eta}$ in Eq.(2): this fixes the value of $\eta$
through the condition:
$$
N={\Gamma (\eta/2-1)\Gamma (2-\eta)\Gamma (D-2+\eta)\Gamma (1-\eta/2)
\over \Gamma (D/2+\eta -2)\Gamma (D/2+\eta/2-1)\Gamma (D/2-\eta/2+1)\Gamma
(2-\eta-D/2)}.
\eqno(5)
$$
This is precisely the result of the SCSA for the $O(2N)$-vector model as
expected.
As long as $D\le 4$, one can check the assumption that $\eta$ leads to $\Pi
>>1$
at low momenta (when $D>4$, then $\Pi$ is negligible and one encounters only
the
Gaussian fixed point).

The other fixed point corresponds to $v\neq 0$: assuming also $\Pi >>1$ we use
the
asymptotic behaviour ${\hat v}(p)\approx 1/N\Pi$ and ${\hat u}(p)\approx
1/2N\Pi$,
deduced from the series (3). These leading terms add up to give a factor of
$3/N$
in the right-hand side of the Dyson equation and thus the only modification
with
respect to the pure Heisenberg case is that in Eq(5) the factor N should be
replaced by N/3:
$$
{N\over 3}={\Gamma (\eta/2-1)\Gamma (2-\eta)\Gamma (D-2+\eta)\Gamma (1-\eta/2)
\over \Gamma (D/2+\eta -2)\Gamma (D/2+\eta/2-1)\Gamma (D/2-\eta/2+1)\Gamma
(2-\eta-D/2)}.
\eqno(6)
$$
This equation leads to a $\eta_{SCSA}$ which is well-behaved in the physical
case
N=2,3, D=3. It can be solved now numerically or expanded in various limits.
Numerically $\eta_{SCSA}(N=3, D=3)\approx 0.11$ and $\eta_{SCSA} (N=2,
D=3)\approx
0.15$, a perfectly sensible result.  Present Monte-Carlo estimates$^3$ favor a
smaller
$\eta$ for N=3 since $\gamma /\nu =2.011(14)$. The SCSA should not be expected
to be quantitative: this is the mere existence of the fixed point which is the
relevant information.

The large-N limit of the SCSA agrees by construction with the result of the
direct  large-N result$^{12}$:
one recovers the result $\eta =6[4/D-1]S_D/N$ where $S_D =\sin [\pi (D-2)/2]
\Gamma (D-1)/(2\pi \Gamma^2 (D/2))$. In the neighborhood of D=4, one finds
$\eta_{SCSA}=3\epsilon^2 /4N+O(\epsilon^3)$. This $\epsilon$-expansion is
well-behaved for any N: in the SCSA there is no hint of a fluctuation-induced
first-order transition. This limiting case matches the $\epsilon$-expansion
only if one takes also the limit $N\rightarrow\infty$. In fact,
$\eta_{4-\epsilon}=
f(N)\epsilon^2 +O(\epsilon^3)$ with $f(N)$ a complicated function of N which
has
the asymptotic behaviour $f(N)\approx 3/4N$ as $N\rightarrow\infty$.

In the case of the NS transition, Radzihovsky has shown$^8$ that $\eta_{SCSA}$
is
well-behaved near D=4 but has a singularity in the $\epsilon$-expansion. This
has
led him to propose that the lack of stable fixed point seen in the
$\epsilon$-expansion studies$^7$ is due to a breakdown of the expansion itself
instead of a fluctuation-induced first-order phase transition.
Since it is difficult to estimate the validity of the SCSA itself, it may be
that
the SCSA is not accurate enough to capture the fluctuation-induced transition.
In the present
context of noncollinear magnets, the situation is slightly different:
$\eta_{SCSA}$
is well-behaved near D=4 {\it and} has a regular $\epsilon$-expansion for any
N.

In the neighborhood of the lower critical dimension, the exponent $\eta$ can be
expanded in D=2+$\epsilon$ with the result $\eta_{SCSA}= 3\epsilon /2(N-3)
+O(\epsilon^2)$. There is a singularity at N=3 in the $\epsilon$-expansion due
to the fact that $\eta$ no longer vanishes (in the SCSA) at D=2 when $N<3$.
For N=3, one has from the SCSA (equation 6) $\eta$=0 at D=2 which is coherent
with
the fact that there is no phase transition.
This behaviour
appears of course in the SCSA treatment of the O(N)-vector model since Eqs 5
and 6
differ by the substitution $N\rightarrow N/3$. However in the O(N) case,
$\eta_{SCSA}=\epsilon/(N-2)$ near D=2 and $\eta_{SCSA}$ is nonzero at D=2 when
$N<2$. Since we expect in a N-vector model that in D=2 $\eta$ vanishes for
$N>2$,
this means that the N-dependence from SCSA cannot be blindly believed near N=3.

Noncollinear magnets have been studied near two dimensions in a sigma model
approach$^{11}$. The symmetry breaking pattern defines a homogeneous
non-symmetric
manifold $G/H=O(N)\times O(2)/O(N-2)\times O(2)_{diag}$ that specifies uniquely
a
nonlinear sigma model. For any finite N there is a fixed point in
D=2+$\epsilon$
which merges smoothly with the fixed point obtained in the large-N limit of the
linear theory. The exponent $\eta$ is given by$^{14,11}$:
$$
\eta_{2+\epsilon}= {3N^2-10N+9\over 2(N-2)^3}\epsilon +O(\epsilon^2).
\eqno(7)
$$
For large N, this agrees with $\eta_{SCSA}\approx 3\epsilon /2N$. In addition
the
formula (7) has the correct feature that it blows up for N=2 instead of N=3 as
found in the SCSA. For N=3, the sigma model has a peculiarity: since
$O(3)\times
O(3)\equiv O(4)$, it describes O(4) critical behaviour. This is incompatible
with
numerical simulations$^3$ in D=3. The reason for this failure is not known at
the
present time. A view has been reported$^{14}$ that the inability of the
$\epsilon
=D-2$ expansion to detect the nontrivial topological structure of the
order-parameter space might be the reason for this failure.
It is clear that the SCSA cannot shed any light on
this problem since its functional dependence upon N is too approximate. In the
SCSA there is no hint of any symmetry enhancement since O(4) critical behaviour
cannot be reached by the linear theory (1).

In this Letter, I have studied the phase transition occuring in noncollinear
magnets that break the full rotation group at low temperatures. A
self-consistent
screening approximation valid for all dimensions and any number of components
of
the order parameter leads  to a second-order phase transition in the physical
case
of Heisenberg and XY systems in D=3. There is no sign of fluctuation-induced
first-order phase transition contrary to the result of the standard
$4-\epsilon$-expansion. This is in agreement with Monte-Carlo results on model
systems that possess the correct symmetry-breaking pattern. There are also some
experimental results$^{11}$ that may be explained by the existence of the
corresponding universality class. It is worth pointing out that the SCSA
approximation is not a controlled approximation with a systematic expansion
parameter and thus it is difficult to estimate its validity.
The whole picture is close to the
normal-to-superconducting phase transition as studied by Radzihovsky: here the
SCSA leads also to a second order transition in the whole (N, D)-plane instead
of
the first-order transition predicted by the $4-\epsilon$-expansion. There is
however a difference: in the NS case $\eta_{SCSA}$ is well-behaved but has a
singularity when expanded in $\epsilon$ while in helical magnets even the
expansion is regular. In the two cases (NS and helical magnets) the transition
is
continuous near two dimensions$^{11,15}$ and in D=3: this is reproduced by the
SCSA calculation.

\bigskip
\bigskip
\noindent {\bf Acknowledgments}: I thank F. David, E. Guitter and J.
Zinn-Justin
for discussions.
\vfill
\eject
\centerline{\bf REFERENCES}
\bigskip
\item{[1]}For a review, see M. L. Plumer, A. Caill\'e, A. Mailhot and H. T.
Diep in
``Magnetic Systems with Competing Interactions'', H. T. Diep editor, World
Scientific, Singapore 1994.
\medskip
\item{[2]}H. Kawamura, J. Phys. Soc. Jpn. {\bf 61}, 1299 (1992).
\medskip
\item{[3]}T. Bhattacharya, A. Billoire, R. Lacaze and Th. Jolicoeur, J. Phys. I
(Paris) {\bf 4}, 181 (1994).
\medskip
\item{[4]}D. Loison and H. T. Diep, Phys. Rev. B{\bf 50}, 16453 (1994).
\medskip
\item{[5]}H. T. Diep, Phys. Rev. B{\bf 39}, 3973 (1989).
\medskip
\item{[6]}T. Garel and P. Pfeuty, J. Phys. C{\bf 9}, L245 (1976); see also D.
Bailin, A. Love and M. A. Moore, J. Phys. C{\bf 10}, 1159 (1977).
\medskip
\item{[7]}B. I. Halperin, T. C. Lubensky and S. K. Ma, Phys. Rev. Lett. {\bf
32},
292 (1974).
\medskip
\item{[8]}L. Radzihovsky, Europhys. Lett. {\bf 29}, 227 (1995).
\medskip
\item{[9]}A. J. Bray, Phys. Rev. Lett. {\bf 32}, 1413 (1974).
\medskip
\item{[10]}C. Dasgupta and B. I. Halperin, Phys. Rev. Lett. {\bf 47}, 1556
(1981).
\medskip
\item{[11]}P. Azaria, B. Delamotte and Th. Jolic\oe ur, Phys. Rev. lett. {\bf
26},
3175 (1990); P. Azaria, B. Delamotte, F. Delduc and Th. Jolic\oe ur, Nucl.
Phys.
B{\bf 408}, 485 (1993).
\medskip
\item{[12]}H. Kawamura, Phys. Rev. B{\bf 38}, 4916 (1988).
\medskip
\item{[13]}P. Le Doussal and L. Radzihovsky, Phys. Rev. Lett. {\bf 69}, 1209
(1992).
\medskip
\item{[14]}H. Kawamura, J. Phys. Soc. Jpn. {\bf 60}, 1839 (1991).
\medskip
\item{[15]}I. D. Lawrie, Nucl. Phys. B{\bf 200}[FS4], 1 (1981); I. D. Lawrie
and C.
Athorne, J. Phys. A{\bf 16}, 4587 (1983).
\medskip
\vfill
\eject
\nopagenumbers

\hfuzz=5pt
\baselineskip 12pt plus 2pt minus 2pt
\centerline{\bf A SELF-CONSISTENT THEORY OF PHASE TRANSITIONS}
\centerline{\bf IN NONCOLLINEAR MAGNETS}
\vskip 24pt
\centerline{Th. Jolic\oe ur\footnote{*}{C.N.R.S. Research Fellow,
thierry@amoco.saclay.cea.fr}}
\vskip 12pt
\centerline{\it Service de Physique Th\'eorique}
\centerline{\it C.E.  Saclay}
\centerline{\it F-91191 Gif-sur-Yvette CEDEX, France}
\vskip 48pt
\vskip 24pt
\vskip 1.0in
\centerline{To appear in: {\it Europhysics Letters}}
\vskip 3.2in
\noindent January 1995

\noindent PACS No: 75.40.C, 75.10.H. \hfill SPhT/95-005
\vfill
\eject
\bye